\documentclass[a4paper]{panl}
\usepackage{cite}
\usepackage{wrapfig}
\usepackage{graphicx}
\usepackage{amssymb}
\usepackage{amsfonts}
\usepackage{amsmath}
\usepackage{longtable}
\usepackage{rotating}
\usepackage{lscape}
\usepackage{epsfig}
\usepackage{multirow}

\usepackage{lineno}

\originalTeX
\begin{document}

\title{\bf On transverse single-spin asymmetries in $D$-meson production at the SPD NICA experiment
}

\maketitle
\authors{A.\,V.\,Karpishkov$^{a,b,}$\footnote{E-mail: karpishkoff@gmail.com},
V.\,A.\,Saleev$^{a,b,}$\footnote{E-mail: saleev@samsu.ru}}
\setcounter{footnote}{0} \from{$^{a}$\,Samara National Research
University, Moskovskoe Shosse, 34, 443086, Samara, Russia}
\from{$^{b}$\,Joint Institute for Nuclear Research, Joliot-Curie St,
6, 141980, Dubna, Russia}


\begin{abstract}
In the present study we are interested in the Sivers effect in
process $p^{\uparrow}p\to DX$. We calculate the transverse
single-spin asymmetry (TSSA) of $D$-meson production within two
phenomenological models, namely the Generalized Parton Model (GPM)
and it's Colour Gauge-Invariant formulation (CGI-GPM), which takes
into account the transverse momenta of initial partons. The last one
allows us to study process-independent Sivers functions. To predict
production cross-section of $D$-mesons we use a fragmentation
approach with scale-independent Peterson fragmentation function,
taking in mind nonzero masses of $c$-quark and $D$-mesons. Estimates
for the TSSAs, as a function of transverse momentum and Feynman
variable $x_F$ of $D$-meson, within the conditions of the planned
SPD NICA experiment are presented for the first time.\end{abstract}
\vspace*{6pt}

\label{sec:intro}
\section*{\bf Introduction}
To the present time a problem of description of a proton
three-dimensional structure and its spin properties remain unsolved.
The Transverse Momentum Dependent (TMD) Parton Distribution
Functions (PDFs) can allow us access to that
information~\cite{Collins:1989gx,Angeles-Martinez:2015sea}. One of
the widely used instruments for investigation of proton spin
structure in collisions $p^{\uparrow}p \to h X$ of polarized protons
with the unpolarized ones is a Sivers
effect~\cite{Sivers:1989cc,Boer:1997bw}. A Sivers function provides
access to the density of unpolarized gluons $g$ (or quarks $q$) with
intrinsic transverse-momentum ${\bf q}_T$ inside a transversely
polarized proton $p^{\uparrow}$, with three-momentum $\bf P$ and
spin polarization vector~$\bf S$,
\begin{equation}
F^{\uparrow}_g(x,{\bf q}_T)=F_g(x,q_T)+\frac{1}{2}\Delta^N
F_g^{\uparrow}(x,q_T){\bf S}\cdot (\hat{\bf P}\times \hat {\bf
q}_T),
\end{equation}
where $x$ is the proton light-cone momentum fraction carried by the
gluon, $F_g(x,q_T)$ is theunpolarised TMD parton density, $\Delta^N
F_g^{\uparrow}(x,q_T)$ is the Sivers function, $q_T=|{\bf q}_T|$ and
symbol $(\ \hat {}\ )$ denotes a unit vector, $\hat {\bf a}={\bf
a}/|{\bf a}|$.

In the present work we study the Sivers effect in the production of
$D$-mesons at the NICA collider, which in perspective can allow us
to refine parameters of a gluon Sivers function (GSF) through
measurements of the~TSSA.

\label{sec:theory}
\section*{\bf Theoretical models}

When we are interested in the production of final-state particles
with transverse momenta much smaller than the hard scale of the
reaction $\mathbf{k}_T\ll\mu$ (which is set by a heavy final state),
we already can't neglect small momenta of initial-state particles
$\sqrt{\langle\mathbf{q}_T^2\rangle}\simeq|\mathbf{k}_T|$. So the
collinear parton model (CPM) is no longer applicable in that region.
The TMD-factorization is proven in the limit of small transverse
momenta of final-state particles, so it is more relevant in the
region $\mathbf{k}_T\ll\mu$~\cite{Collins:2011zzd}. For the
phenomenological purposes a Generalized Parton Model (GPM) can be
applied as a simplified version of TMD-factorization even in
processes for which TMD factorization has not been rigorously proven
yet.

The main idea of the GPM is that the TMD PDFs can be parametrized by a simple factorized ansatz:
\begin{equation}
F_a(x,q_T,\mu_F)=f_a(x,\mu_F)G_a(q_T),
\end{equation}
where $f_a(x,\mu_F)$ is corresponding collinear PDF for parton
"\emph{a}", and the dependence on transverse parton momentum is
described by a Gaussian distribution $G_a(q_T)= \exp[{-q_T^2/\langle
q_T^2\rangle_a}]/(\pi\langle q_T^2\rangle_a)$ with normalization
condition $\int d^2q_T G_a(q_T)=1$.

Such a way, a master-formula for the differential cross section of
production of heavy final state $\mathcal{C}$ in collision of two
protons can be written as:
\begin{multline}
d\sigma (pp\to {\cal C}X)=\int dx_1 \int d^2{\bf q}_{1T} \int dx_2 \int
d^2{\bf q}_{2T} \times\\
\times F_a(x_1,{\bf q}_{1T},\mu_F) F_b(x_2,{\bf q}_{2T},\mu_F)
d\hat\sigma\label{eq:master},
\end{multline}
where $d\hat{\sigma}$ is the partonic cross section for $ab\to {\cal C}X$ partonic subprocess.

The TSSA under study in the present paper for the process $p^{\uparrow}p\to DX$ has the following standard definition:
\begin{equation}\label{eq:TSSAdef}
A_N=\frac{d\sigma^{\uparrow}
-d\sigma^{\downarrow}}{d\sigma^{\uparrow}
+d\sigma^{\downarrow}}=\frac{d \Delta\sigma}{2d\sigma}.
\end{equation}

The numerator and denominator of $A_N$ have the form:
\begin{multline}\label{eq:TSSAnum}
\!\!\!\!d\Delta\sigma\propto\!\int\!dx_1\!\int\!d^2q_{1T}\!\int\!dx_2\!\int\!
d^2q_{2T}\!\int\! dz\bigl[\hat F_g^{\uparrow}(x_1,{\bf q}_{1T},\mu_F)-\hat
F_a^{\downarrow}(x_1,{\bf q}_{1T}, \mu_F) \bigr]\times\\
\times F_b(x_2,q_{2T},\mu_F) d\hat\sigma(ab\to c\bar cX)D_{D/c}(z,\mu_F),
\end{multline}
\begin{multline}\label{eq:TSSAden}
\!\!d\sigma\propto\!\int\!dx_1\!\int\!d^2q_{1T}\!\int\!dx_2\!\int\!d^2q_{2T}\!\int\! dz
F_a(x_1,q_{1T},\mu_F) F_b(x_2,q_{2T},\mu_F)\times\\
\times d\hat\sigma(ab\to c\bar cX)D_{D/c}(z,\mu_F),
\end{multline}
where $\hat F_a^{\uparrow,\downarrow}(x,q_{T},\mu_F)$ is the
distribution of unpolarized parton $a$ in polarized proton and
$D_{D/c}(z,\mu_F)$ is $D$-meson fragmentation function~(FF) with
light-cone momentum fraction is defined as
$z=p_D^+/k_c^+$~\cite{DAlesio:2004}. Here as a hadronization model
we use a fragmentation approach, which involves a mass of a charm
quark~\cite{DAlesio:2004}. To describe a fragmentation of the charm
quark we use a phenomenological Peterson's FF~\cite{Peterson:1983}
with parameter $\epsilon_c=0.06$.

Within the standard GPM though the GSF is assumed to be a process
dependent. The problem is solved by involving the effects of
initial-state interaction (ISI) and final-state interaction (FSI)
through an additional gluon
exchange~\cite{Gamberg:2011,DAlesio:2017rzj}. In the case of gluon
Sivers effect, the process-dependent GSF can be presented as a
linear combination of two independent and universal GSFs of $f$-type
($F_{1T}^{g(f)}$) and $d$-type ($F_{1T}^{g(d)}$) corresponding to
two independent ways of combining three gluons into a color singlet.
The coupling of additional ``eikonal'' gluon from the GSF to the
hard process leads only to modification of the color structure of
the latter one.

A similar study we have already performed for a production of
another heavy final state -- $J/\psi$ mesons~\cite{Karpishkov:2021}.
But charmonia production is not the only channel for the GSF
measurements at the NICA collider, and in the following section we
provide a predictions of the $A_N$ for the $D$-meson production as
another probe of the GSF.

\label{sec:results}
\section*{\bf Numerical predictions for the NICA facility}

Here we present our theoretical results for the TSSA $A_N^D$ of the
$D$-meson production within the kinematics of the SPD NICA
experiment. These kinematic conditions define the total collision
energy as $\sqrt{s}=27$~GeV, the rapidity of produced $D$-mesons
must be in the interval $|y_D|<3$ and the transverse momentum of
$D$-meson $0<p_T<3$~GeV. We calculate the TSSA as a function of the
transverse momentum $p_T$ and the Feynman variable $x_F$ of the
produced $D$-meson. In all our calculations we take into account a
charm quark mass equal to $m_c=1.2$~GeV.

The renormalization and factorization scales we define to be equal
to $\mu_F=\mu_R=\xi \sqrt{k_T^2+m_c^2}$, where $k_T$ is the
transverse momentum of a charm quark. To estimate theoretical
uncertainties, which come from the arbitrariness in a choice of the
hard scale, we vary the parameter $\xi$ in a closed interval
$1/2\leq\xi\leq2$. The uncertainties are shown on our plots as
shaded bands.

In the Fig.~\ref{fig01} we compare the predictions of TSSA within
the GPM in two different parametrizations of the GSF, namely a
parametrization of D'Alesio with co- uthors~\cite{DAlesio:2018rnv}
(red solid line) and the SIDIS1
parametrization~\cite{DAlesio:2015fwo} (blue dashed line). We note
that TSSA in SIDIS1 parametrization is significantly larger than one
in D'Alesio parametrization.

\begin{figure}[h!]
\begin{center}
\includegraphics[width=0.5\textwidth]{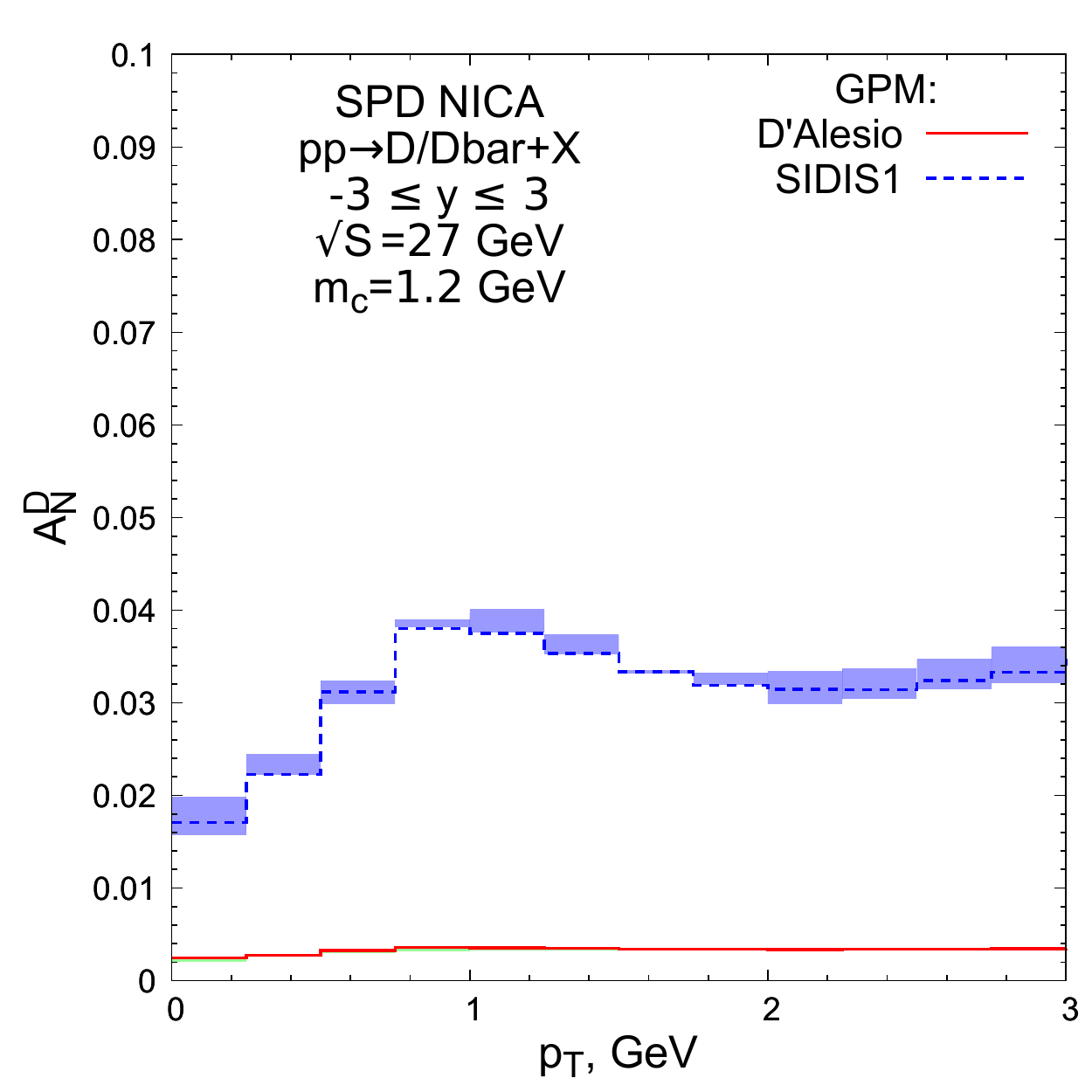}\includegraphics[width=0.5\textwidth]{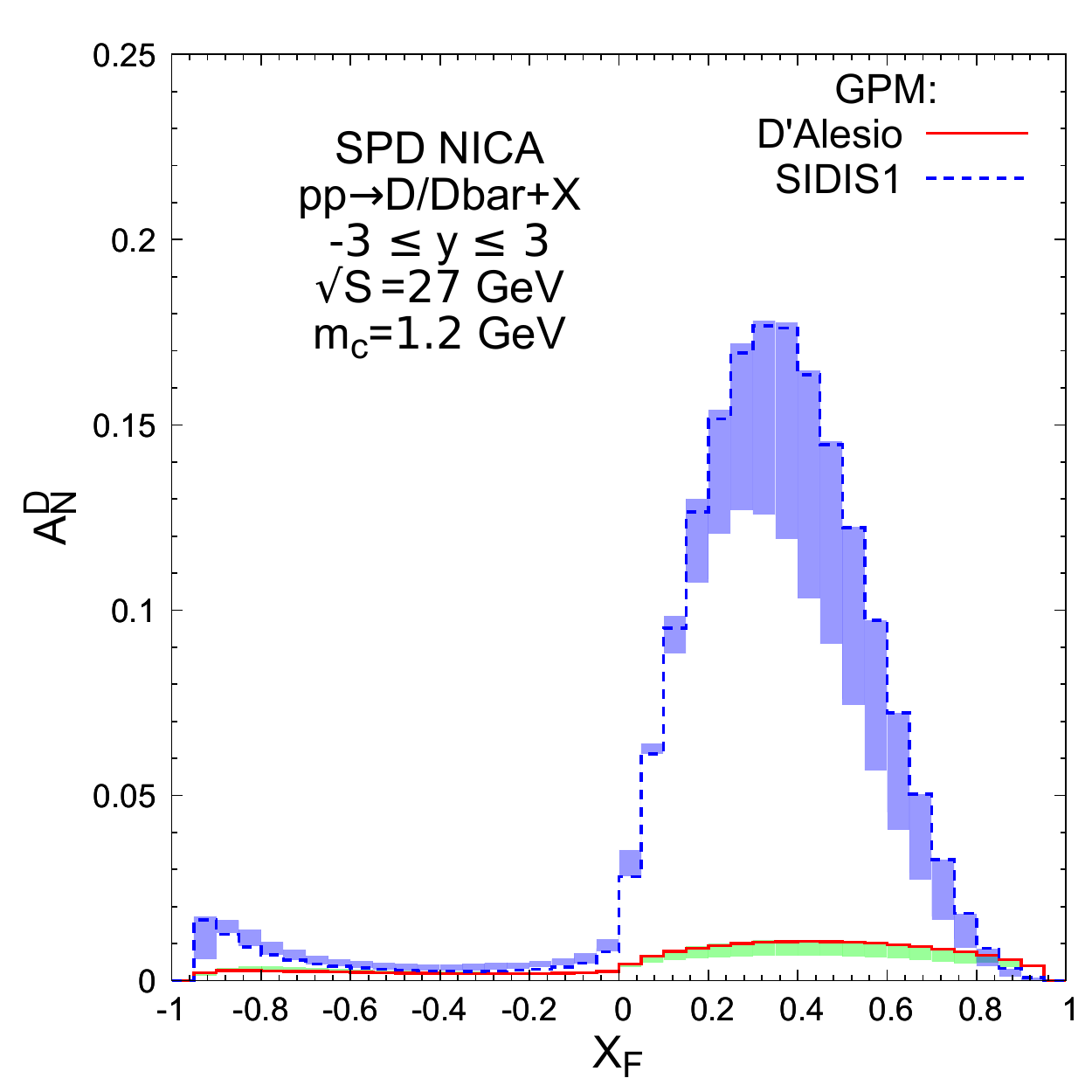}
\vspace{-3mm}
\caption{Predictions for TSSA on SPD NICA within the GPM as a function of transverse momentum (\emph{left} panel) and $x_F$ (\emph{right} panel)  of $D$-meson. The TSSA in D'Alesio parametrization is shown by a \emph{red solid} line, SIDIS1 -- by a \emph{blue dashed} line.}
\end{center}
\labelf{fig01}
\vspace{-5mm}
\end{figure}
\begin{figure}[h!]
\begin{center}
\includegraphics[width=0.5\textwidth]{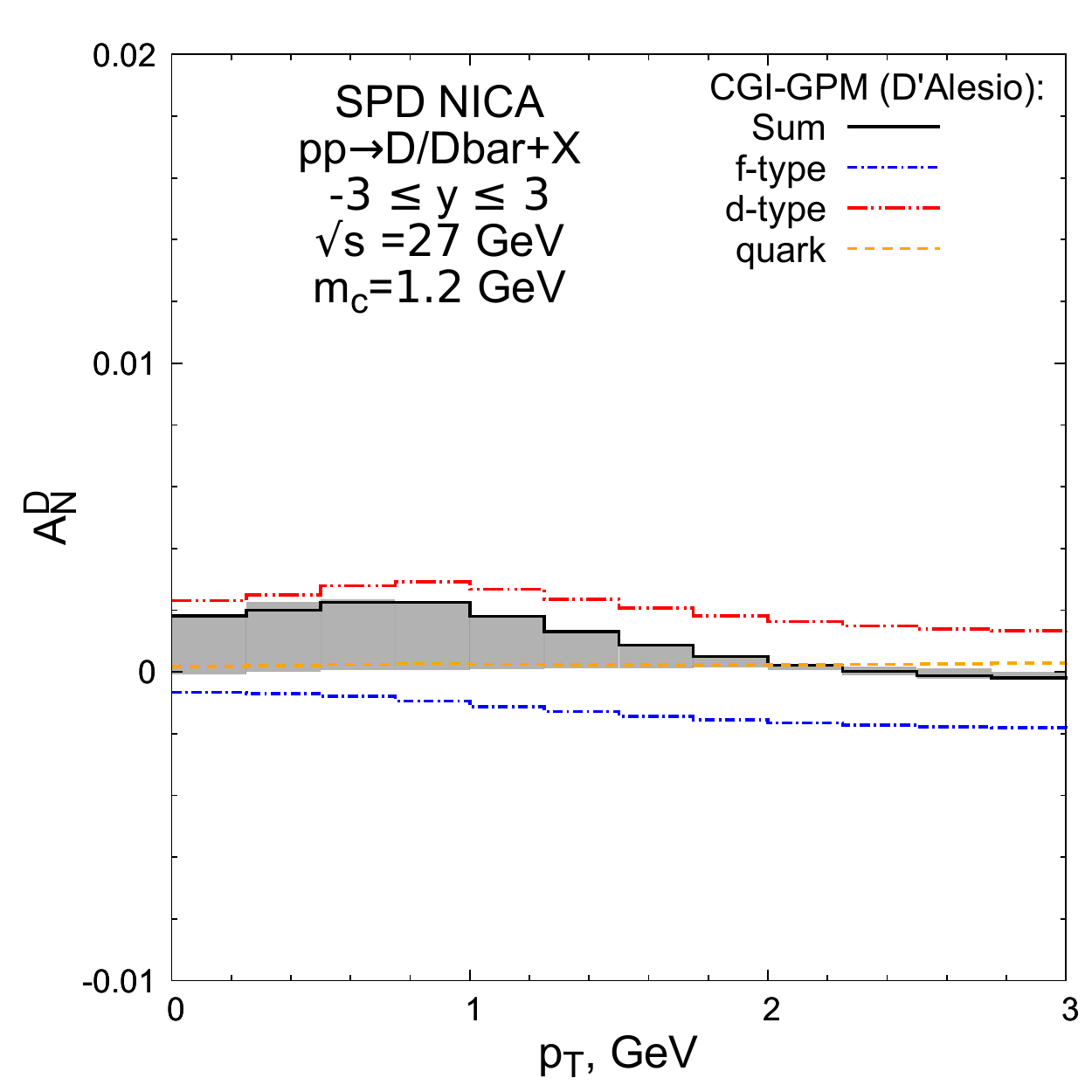}\includegraphics[width=0.5\textwidth]{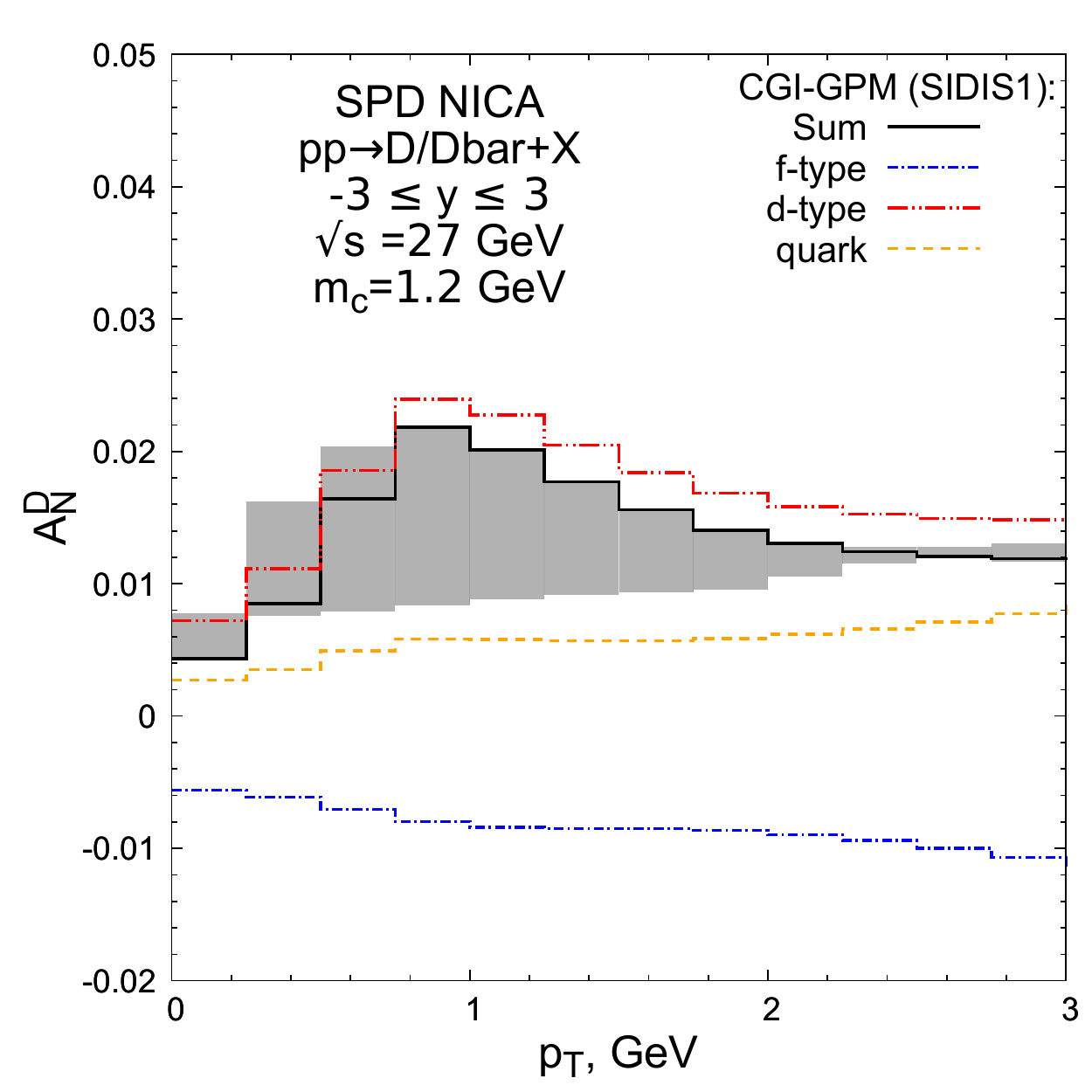}
\vspace{-3mm}
\caption{Predictions for TSSA on SPD NICA within the CGI-GPM as a function of $p_T$ of $D$-meson with the D'Alesio parametrization (\emph{left} panel) and SIDIS1 parametrization (\emph{right} panel) of the GSF. Orange dashed lines stands for contribution of the quark Sivers function, blue dash-dotted lines -- for contribution of the $f$-type GSF, and red double-dot-dashed lines -- for the $d$-type GSF.}
\end{center}
\labelf{fig02}
\vspace{-5mm}
\end{figure}

The predictions of the TSSA within the CGI-GPM are presented on
Fig.~\ref{fig02} and Fig.~\ref{fig03}. On the first one, the TSSA is
shown as a function of the transverse momentum of $D$-meson, while
on the second one the TSSA is a function of $x_F$. On the left
panels of corresponding figures, we put the TSSA in D'Alesio
parametrization of the GSF, and plots on the right panels correspond
to the TSSA in the SIDIS1 parametrization. By orange dashed lines we
depicted a contribution of a quark Sivers function to the TSSA. Blue
dash-dotted lines correspond to the $f$-type of the GSF, while red
double-dot-dashed lines correspond to the $d$-type of the GSF. Black
solid lines are the sum of all the contributions, and shaded bands
show summary uncertainties.

\begin{figure}[h!]
\begin{center}
\includegraphics[width=0.5\textwidth]{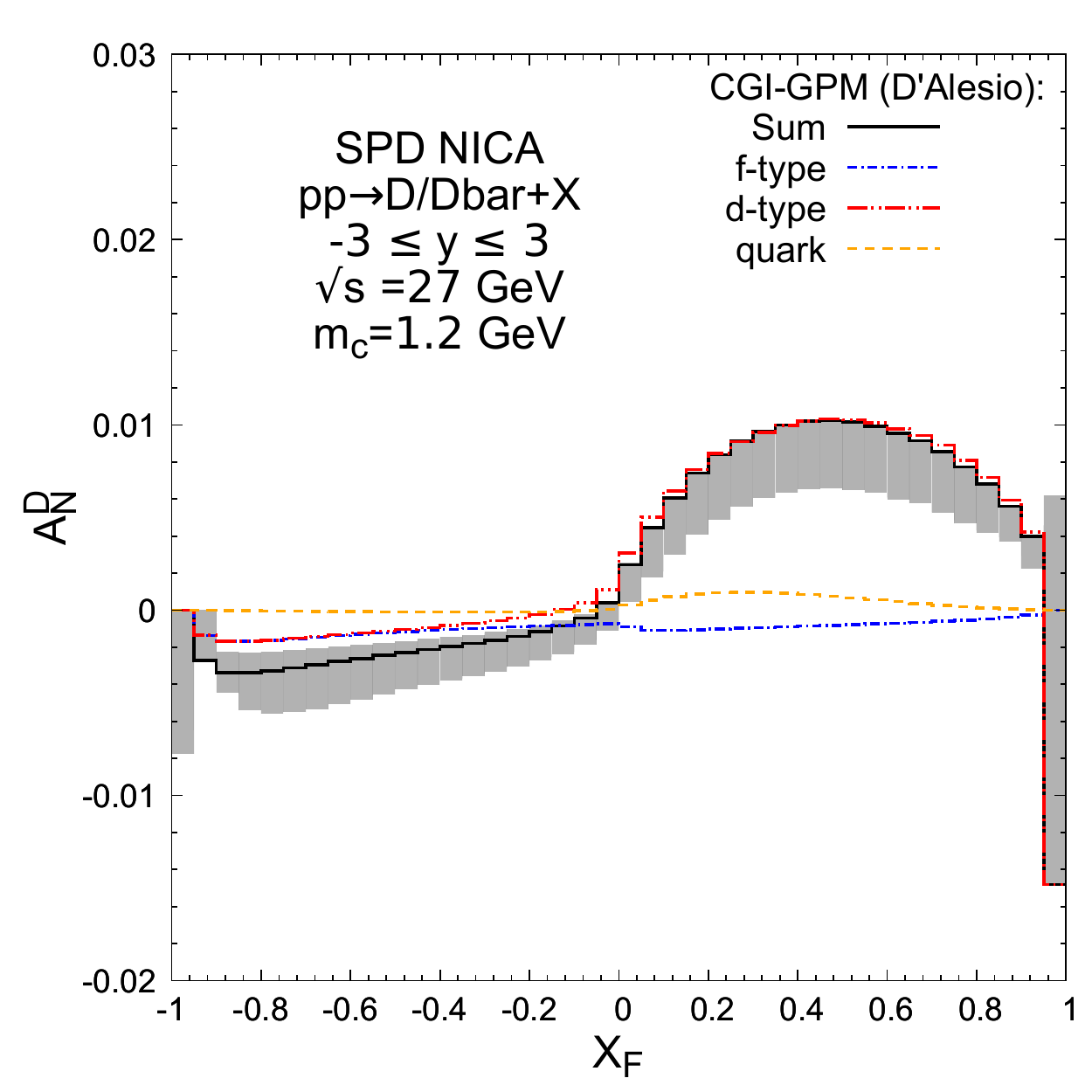}\includegraphics[width=0.5\textwidth]{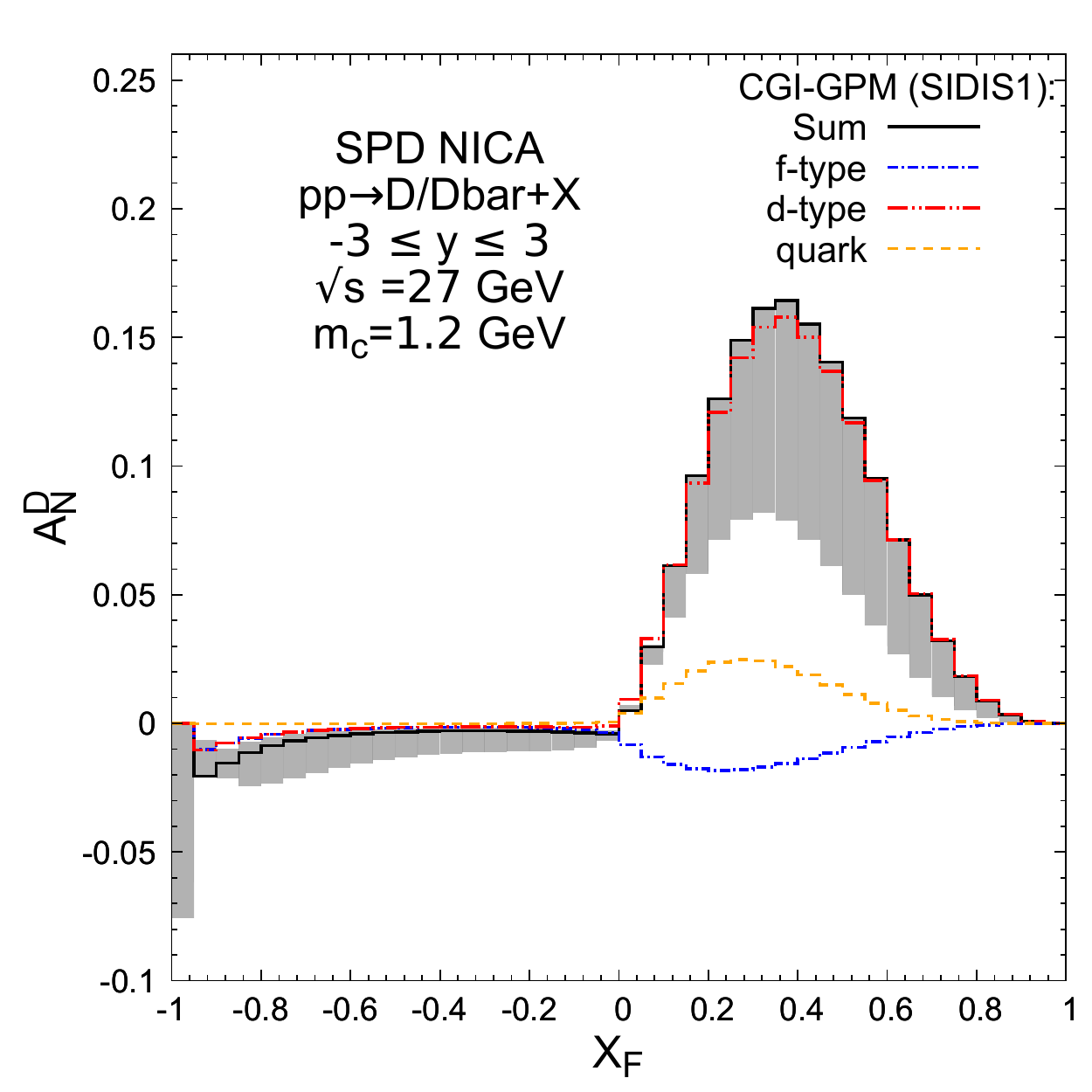}
\vspace{-3mm}
\caption{Predictions for TSSA on SPD NICA within the CGI-GPM as a function of $x_F$ of $D$-meson with the D'Alesio parametrization (\emph{left} panel) and SIDIS1 parametrization (\emph{right} panel) of the GSF. The notations are the same as in Fig.~\ref{fig02}.}
\end{center}
\labelf{fig03}
\vspace{-5mm}
\end{figure}

The predictions of the CGI-GPM allow negative values for the TSSA,
while the GPM predicts strictly positive ones. Moreover, we should
note that the $d$-type GSF within the CGI-GPM is the dominant one,
and the other two contributions almost cancel each other. We see
also that as the GPM, as the CGI-GPM show that the TSSA in D'Alesio
parametrization of the GSF is much smaller than the one in SIDIS1.
Precise experimental measurements can in principle allow us to
choose a more relevant parametrization of the GSF within both the
GPM and the CGI-GPM.

\newpage
\bibliographystyle{pepan}
\bibliography{biblio}

\end{document}